\documentclass[reprint,superscriptaddress,nofootinbib,twocolumn,amsmath,amssymb,aps,pra,longbibliography]{revtex4-1}

\usepackage{graphicx}% Include figure files
\usepackage{dcolumn}% Align table columns on decimal point
\usepackage{bm}% bold math
\usepackage[colorlinks]{hyperref}

\begin{document}

\preprint{APS/123-QED}

\title{Design of mid-infrared entangled photon sources using lithium niobate}% Force line breaks with \\

\author{Jin-Long Zhu}
\altaffiliation{These authors contributed equally to this work.}
\author{Wen-Xin Zhu}
\altaffiliation{These authors contributed equally to this work.}
\author{Xiao-Tao Shi}
\altaffiliation{These authors contributed equally to this work.}
%\address{\authormark{*} These authors contributed equally to this work.}
\author{Chen-Tao Zhang}
\author{Xiangying Hao}
\email{xyhao.321@163.com}
\author{Zi-Xiang Yang}
\email{yangzixiangyzx@foxmail.com}
\author{Rui-Bo Jin}
\email{jin@wit.edu.cn}

\address{Hubei Key Laboratory of Optical Information and  Pattern Recognition, Wuhan Institute of Technology, Wuhan 430205, PR China\\}


\begin{abstract}
The mid-infrared (MIR) band entangled photon source is vital for the next generation of quantum communication, quantum imaging, and quantum sensing. However, the current entangled states are mainly prepared in visible or near-infrared bands. It is still  lack of high-quality entangled photon sources in the MIR band. In this work, we optimize the poling sequence of lithium niobate to prepare two kinds of typical entangled states, the Hermit-Gaussian state and the comb-like entangled state at 3.2 $\mu$m. We have also calculated  the photon pair rates and estimated the effect of fabrication resolution in the schemes. Our approach will provide entangled photon sources with excellent performance for the study of quantum information in the MIR band. 

\end{abstract}

\maketitle

%\tableofcontents

\section{\label{sec:level1}Introduction}

The entangled photon source in the mid-infrared (MIR) band (approximately 2-20 $\mu$m) is promising for the next generation of quantum communication, quantum  imaging, and quantum sensing\cite{Tournie2019book, Ebrahim-Zadeh2008book}. 
In  quantum communication, the entangled photon source in the wavelength between 3 µm and 5 µm covers the atmospheric transmission window, which has higher transparency than that in the near-infrared band  and is beneficial for free-space quantum communications \cite{Bellei2016oe}.
In quantum imaging, room temperature objects emit light at MIR wavelengths, therefore the MIR band entangled photon sources are compatible with novel applications in infrared thermal imaging \cite{Tittl2015, Mancinelli2017}.
In  quantum sensing, the MIR band entangled photon source has  strong absorption bands of a variety of gases, which leads to essential applications in gas quantum  sensing \cite{Shamy2020}. With the help of entanglement, the sensing precision may be improved from the  shot noise limit to the Heisenberg limit \cite{zhiyuanzhou}.

Spontaneous parametric down-conversion (SPDC) is one of the widely used methods to prepare an entangled photon source. Recently, several works have investigated the generation of  entangled photons in MIR range from an SPDC process in the nonlinear crystal. %
From the theoretical side,  
in 2016, Lee et al reported a scheme for the generation of polarization-entangled state from periodically poled potassium niobate (PPKN) covering 3.2 to 4.8 $\mu$m  \cite{Lee2016};
In 2018,  McCracken et al numerically investigated six novel nonlinear crsystals %(PPLN, PPKTP, GaP, GaAs, CdSiP$_2$, and ZnGeP$_2$) 
in order to generate MIR single photons  \cite{McCracken2018}; 
In 2020, Kundys et al numerically studied the reconfigurable MIR single-photon sources in PMN-0.38PT crystal at 5.6 $\mu$m \cite{Kundys2020};
In 2021, Wei et al  theoretically investigated the preparation of MIR spectrally uncorrelated biphotons from an SPDC process using doped lithium niobate (LN) crystals \cite{Wei2021};
These schemes for single photon can also be upgraded to prepare entangled photons.
From the experimental side, in 2020, Prabhakar et al demonstrated an entangled photons source at 2.1 $\mu$m generated from type-II PPLN crystals \cite{Prabhakar2020}.

PPLN is one of the most promising crystals for MIR entangled photon source, not only for its large nonlinear coefficient and wide transparency range \cite{Nikogosyan2005, Liu2017}, but also for its group velocity-matched (GVM) wavelengths, which are in the MIR range \cite{Wei2021}.
However, from the viewpoint of quantum state engineering, the previous entangled source with PPLN is still not optimal \cite{Prabhakar2020,Wei2021}. 
 For a standard PPLN crystal, the phase matching function (PMF) has a ``sinc'' distribution, which has side lobes and will harm the spectral purity of heralded single photon.
To overcome the problem of side lobes, one useful method is to adopt the ``customized poling'' instead of ``periodical poling''\cite{Graffitti2017QST, Drago2022, Morrison2022, Pickston2021}.

Many previous works have been devoted to the optimization of a poling period in a periodically poled potassium titanyl phosphate (PPKTP) crystal at 1550 nm, and the optimization can be divided into three categories: (1) the optimization of poling order: in 2011, Branczyk et al proposed and experimentally demonstrated the first optimization design of KTP by arranging the poling order \cite{Branczyk2011}, and this approach was further improved by Kaneda et al in 2021 \cite{Kaneda2021}. (2) the optimization of  duty cycle: in 2013, Dixon et al proposed to design the duty cycle of KTP \cite{Dixon2013}, which was verified experimentally in 2017 \cite{Chen2017}; In 2019, Cui et al adopted the Adam algorithm in a machine learning framework to optimize the duty cycle \cite{Cui2019PRAppl}. In 2022, Cai et al optimized the duty cycle of LN using the particle swarm algorithm \cite{Cai2022};  (3) the optimization of domain sequence: in 2016, Tambasco et al optimized the domain sequence in the unit of dual domain blocks \cite{Tambasco2016}; In the same year, Dosseva et al proposed to optimize  the sequence of single domain blocks using simulated annealing algorithm \cite{Dosseva2016}; In 2017, Graffitti et al theoretically optimized the domain blocks with sub-coherence-length \cite{Graffitti2017QST}, and then verified experimentally in 2018 \cite{Graffitti2018Optica,Graffitti2020PRL}. %
Recently, frequency-bin entanglement generated by domain-engineered down-conversion has been demonstrated theoretically \cite{Drago2022} and experimentally \cite{Morrison2022,Pickston2021}. These optimization works were mainly focused on PPKTP and at 1550 nm wavelength. However, for the MIR band, PPKTP crystal is no longer applicable because it no longer meets the GVM condition.

In this work, we focus on the MIR band and propose two categories of entangled states at 3.2 $\mu$m, the Hermit-Gaussian (H-G) entangled state and the comb-like entangled state. We will explain how to realize these states by optimizing the poling period of the LN crystal using the domain sequence arrangement algorithm. This work is expected to provide quantum entangled photon sources with excellent performance for the study of quantum information in the MIR band.

\section{Theory}
During the SPDC process, a pump photon with higher frequency impinges on a nonlinear optical crystal, and produces a pair of photons with lower frequency, often referred to as a signal and a idler photon, collectively called  biphoton. 
Without considering higher order nonlinear effects,
the joint spectral amplitude (JSA) of down-converted photons is given by the product of PMF and pump envelope function (PEF)\cite{Mosley2008, Jin2013OE}:
%
%--------------------------------Eq.1-----------------------------------------
\begin{equation}\label{eq1}
f(\omega _s ,\omega _i ) = \phi (\omega _s ,\omega _i ) \times \alpha (\omega _s +\omega _i ),
\end{equation}
where PEF $\alpha$ is determined by the spectral distribution of pump photon and is usually taken as a Gaussian function, and PMF $\phi$ is determined by the properties of the crystal. For periodically poled crystals, PMF can be expressed as a function of $\Delta k$:

%--------------------------------Eq.2-----------------------------------------
\begin{equation} \label{eq2}
\phi (\Delta k) =  \frac{1}{L}\int_0^L {\exp [i\Delta kz]} dz = {\mathop{\rm sinc}\nolimits} (\frac{{\Delta kL}}{2})\exp (\frac{{i\Delta kL}}{2}),
\end{equation}
where $L$ is the crystal length and $\Delta k$ is the difference of the wave vector $ {{\text{k}}_{p(s,i)}}=\frac{{{\omega}_{p(s,i)}}{{n}_{p(s,i)}}}{c}$,
$\Delta k =\frac{{{\omega }_{p}}{{n}_{p}}}{c}-\frac{{{\omega }_{s}}{{n}_{s}}}{c}-\frac{{{\omega }_{i}}{{n}_{i}}}{c}-\frac{2\pi }{\Lambda }$. Using wavelength as the variable, $\Delta k$ can also be written as:
%--------------------------------Eq.3-----------------------------------------
\begin{equation}\label{eq3}
\Delta k = 2\pi  \times (\frac{{n_p (\lambda_p )}}{{\lambda _p }} - \frac{{n_s (\lambda_s )}}{{\lambda _s }} - \frac{{n_i (\lambda_i )}}{{\lambda _i }} - \frac{1}{{\Lambda }}),
\end{equation}
where $n_{p(s,i)}$ and $\lambda_{p(s,i)}$ are respectively the refractive index and the wavelength of the pump (signal, idler) photon. $\Lambda$ is the poling period,
it can be customarily designed according to the target PMF, which will be discussed in Section IV. 
\begin{figure*}[!tbp]
\centerline{\includegraphics[width=18cm]{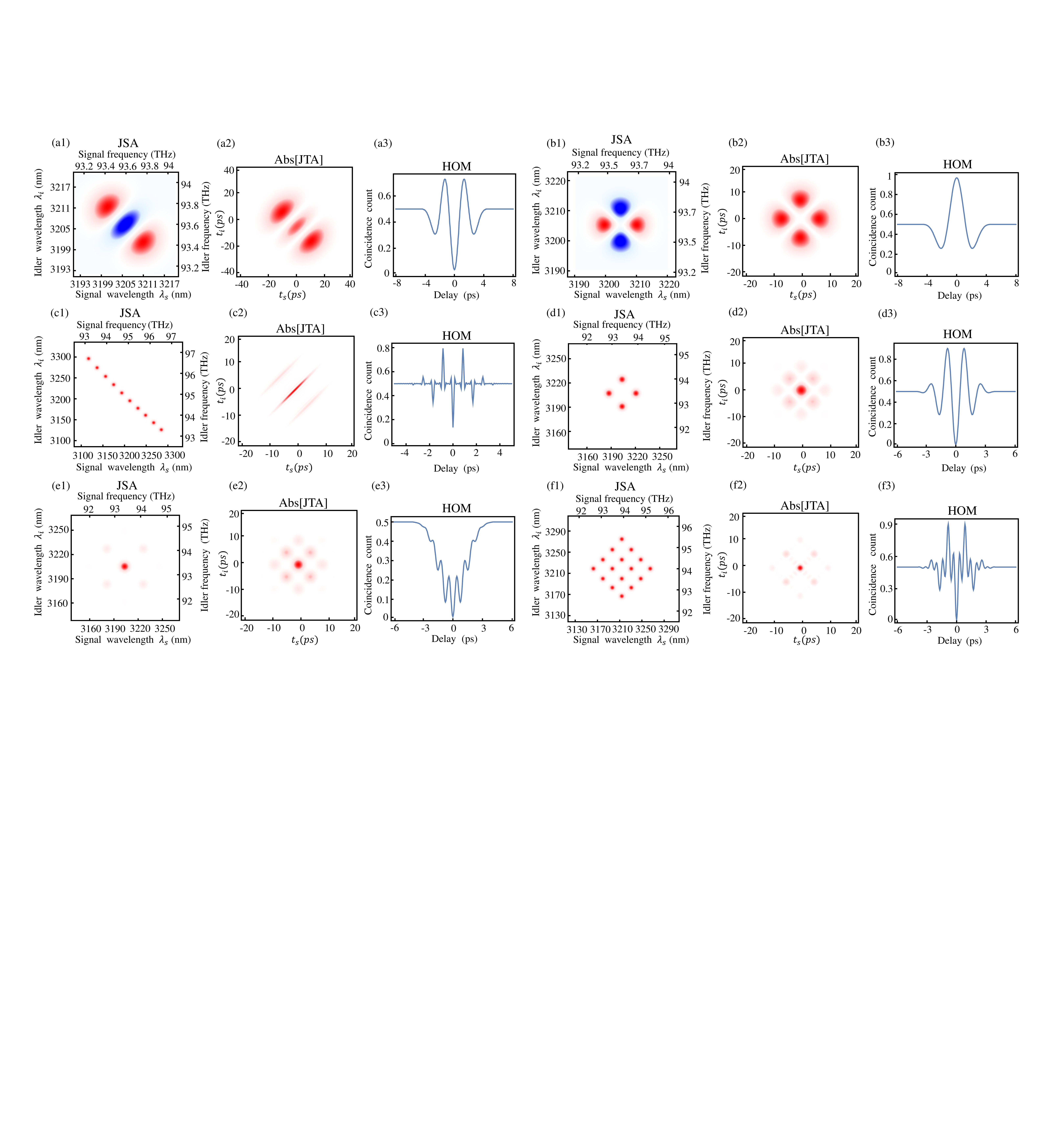}}
\centerline{\parbox[c]{18cm}{\footnotesize Fig.1. \ \
The joint spectral amplitude (JSA), the absolute of the joint temporal amplitude (JTA), and the Hong-Ou-Mandel (HOM) interference patterns for three-mode  Hermit-Gaussian (H-G) state (a1-a3),  four-mode H-G state  (b1-b3), ten-mode comb-like state (c1-c3),  four-mode comb-like state (d1-d3), five-mode comb-like state (e1-e3), and sixteen-mode comb-like state (f1-f3).}} \label{fig1}
\end{figure*}

Due to the energy conservation, $\omega _s +\omega _i=\omega _p$, so the ridge direction of $\alpha (\omega _s +\omega _i ) $ is always distributed along the anti-diagonal direction. To make the $\phi (\omega _s ,\omega _i )$ distributed along the diagonal direction, it is necessary to consider the GVM condition \cite{Jin2013OE}:
%
%--------------------------------Eq.4-----------------------------------------
\begin{equation}\label{eqGVM}
2V_{p}^{-1} = V_{s}^{-1}+V_{i}^{-1},
\end{equation}
where $V_{p(s,i)}^{-1}$ is the inverse of the group velocity of the pump (signal, idler). 
Under the GVM condition, $ \alpha (\omega _s +\omega _i )$ is perpendicular to $\phi (\omega _s ,\omega _i )$, and their product may achieve a single Gaussian mode.
For type-II phase-matched PPLN (o $\rightarrow$ o+e, the pump and the signal is o-ray, and the idler is e-ray), the GVM condition is satisfied at the wavelength of 3207.6 nm \cite{Wei2021} according to the Sellmeier equation \cite{Schlarb1994}. The joint temporal amplitude (JTA) can be obtained by performing an inverse Fourier transform (IFT) on the JSA:
%%--------------------------------Eq.5-----------------------------------------
\begin{equation}\label{eqJSA}
g(t_s, t_i) =  \int_{-\infty}^\infty \int_{-\infty}^\infty f(\omega_s, \omega_i) \exp{(i\omega_s t_s +i\omega_i t_i)} d\omega_s d\omega_i.
\end{equation}
The properties of the JSA and JTA can be measured using the Hong-Ou-Mandel (HOM) interference \cite{Hong1987}, in which the coincidence probability $p$ as a function of time delay $\tau$ can be calculated as \cite{Jin2018Optica}:
%%--------------------------------Eq.6-----------------------------------------
\begin{equation}\label{eqHOMI}
p(\tau) = \frac{1}{2}- \frac{1}{2}\int_{-\infty}^\infty\int_{-\infty}^\infty |f(\omega_s, \omega_i)|^2 \cos {(\omega_s -\omega_i)\tau} d\omega_s d\omega_i.
\end{equation}

\section{Entangled photon sources for MIR wavelengths}
For future applications in quantum communication, quantum imaging, and quantum sensing in the MIR band, we propose two kinds of typical entangled states. The first category is the H-G state, while the second category is the comb-like entangled state. Both categories are intrinsically high-dimensional entangled states and can carry more information for quantum communications \cite{Graffitti2020PRL, Cerf2002}.
Further, the high-dimensional entangled state has a much narrower interference pattern in HOM interference, which is very helpful to improve the precision of a quantum measurement \cite{Lyons2018}.

Figure 1 shows the JSA, absolute of the JTA, and HOM interference patterns for  H-G states and comb-like states.
Figure 1 (a1) shows a typical three-mode JSA, which can be obtained by choosing a PMF of second-order Hermitian function, and a PEF of zero-order Hermitian function.  Figure 1 (a2) is the absolute of the JTA calculated using Eq.(5). 
Figure 1 (a3) is the simulated HOM interference pattern using Eq.(6).
Figure 1 (b1-b3) shows cases of a four-mode H-G state.
Figure 1 (c,d,e,f)  shows cases of ten-mode, four-mode, five-mode, and sixteen-mode comb-like state.
It is interesting to notice in Fig.1 (a1-a2) and (b1-b2) that the H-G mode entangled state has the same mode numbers in both frequency domain and time domain.
In addition, as shown in Fig.1(c3) and (f3), the HOM interference pattern  have a much narrower interference valleys and peaks .

\section{Crystal desgin}
%--------------------HG----------------------------------------------
\subsection{Preparation of H-G entangled state for MIR wavelengths}

In this section, we describe how to prepare the H-G entangled state for MIR wavelengths.
It includes two steps: the first step is to prepare the PMF by designing the LN crystal. The second step is to prepare the PEF by designing the pump laser\cite{ZhangCT}.

%--------------------------------Fig.2-----------------------------------
\begin{figure*}[!tbp]
\centering\includegraphics[width=14cm]{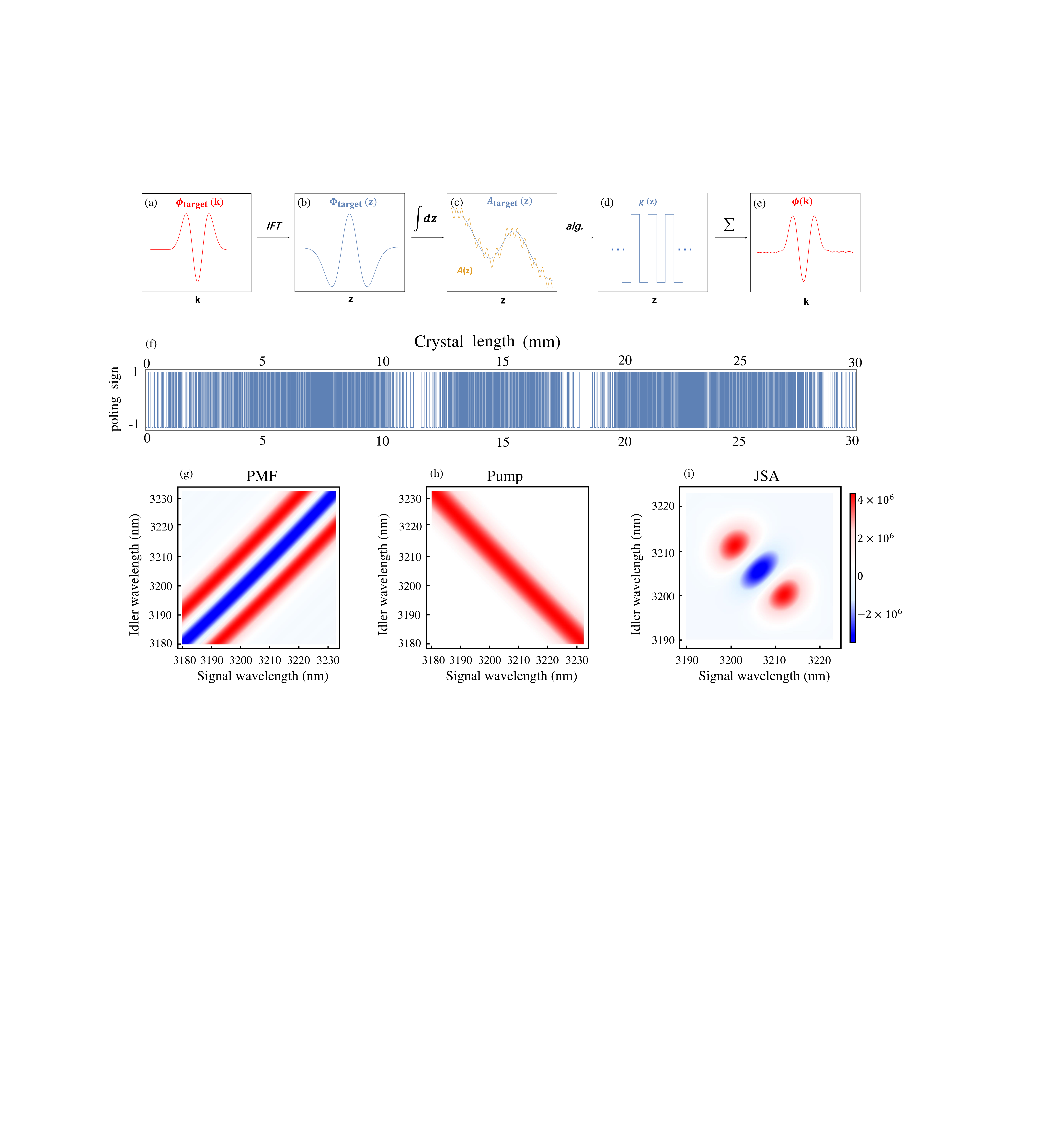}
\centerline{\parbox[c]{13cm}{\footnotesize Fig.2 (a-e) is the design principle for a H-G mode PMF, where ${\phi }_{\text{target}}(k)$ is the designed PMF in wave vector k space, $ \Phi_{\text{target}}(z)$ is the designed PMF in spatial domain, ${A _{\text{target}}}(z)$ is the designed amplitude function of the target light field.  $A(z)$ is the amplitude function of the light field obtained by considering the poling structure in the crystal, $g(z)$ is the poling structure, and  $\phi (k) $ is phase-matching function accumulated in the crystal. (f) is the designed poling distribution of the customized poled lithium niobate (CPLN). (g-i) is the PMF, PEF, and JSA.}} \label{fig2}
\end{figure*}
%--------------------------------Fig.2-----------------------------------

The procedure for designing an H-G mode PMF is shown in Fig.2 (a-e). For the three-mode entangled state, the target PMF, ${\phi }_{\text{target}}(k)$, is the second-order Hermitian function, as shown in Fig.2 (a):
%
%--------------------------------Eq.7-----------------------------------------
\begin{equation} \label{eq7}
\\{{\phi }_{\text{target}}}(k)=\frac{1}{2\sqrt{2}}\exp [-\frac{{{\sigma }^{2}}}{2}{{(k-{{k}_{0}})}^{2}}](-2+4{{(\sigma (k-{{k}_{0}}))}^{2}}),\\
\end{equation}
where ${k}_{0}$ = $2\pi/\Lambda$,  ${k}={{k}_{p}}-{{k}_{s}}-{{k}_{i}}$, $\sigma$ determines the width of the PMF, and we take $\sigma$= $L/6$ . 
According to Eq.(3), we can calculate the poling period $\Lambda$ =  14998.9 nm  when the pump laser is at 1603.8 nm and downconverted photons at 3207.6 nm.
The PMF in the spatial domain, $ \Phi_{\text{target}}(z)$, can be obtained by performing an IFT on Eq.(7):
%
%--------------------------------Eq.8-----------------------------------------
\begin{equation} \label{eq8}
\\\Phi_{\text{target}}(z)=\frac{1}{2\sqrt{\pi }{{\sigma }^{3}}}\exp [-\frac{{{z}^{2}}}{2{{\sigma }^{2}}}](-2{{z}^{2}}+{{\sigma }^{2}})\exp [i{{k}_{0}}z].\\
\end{equation}
The amplitude function of the target light field, ${A _{\text{target}}}(z)$, is obtained by displacing and integrating the amplitude term of $ \frac{1}{2\sqrt{\pi }{{\sigma }^{3}}}\exp [-\frac{{{z}^{2}}}{2{{\sigma }^{2}}}](-2{{z}^{2}}+{{\sigma }^{2}})$ in Eq.(8):

%--------------------------------Eq.9-----------------------------------------
\begin{widetext}
\begin{eqnarray}
A_{\text{target}}(z)&=& C\int\limits_{0}^{z}{\frac{1}{2\sqrt{\pi }{{\sigma }^{3}}}\exp [-\frac{{{({{z}^{'}}-\frac{L}{2})}^{2}}}{2{{\sigma }^{2}}}](-2{{({{z}^{'}}-\frac{L}{2})}^{2}}
+{{\sigma }^{2}}) d{{z}^{'}}}\notag
\\ &=& C\frac{1}{4\sqrt{\pi}\sigma}(2L\text{exp}[-\frac{L^2}{8{\sigma }^2}]+\text{exp}[-\frac{(L-2z)^2}{8{\sigma }^2}] (-2L+4z)\nonumber
 +\sqrt{2\pi}\sigma(-\text{erf}[\frac{L}{2\sqrt{2}\sigma}]+\text{erf}[\frac{L-2z}{2\sqrt{2}\sigma}])).\\
%\\&&
\end{eqnarray}
\end{widetext}
Then we chose proportional coefficient C = ${2\sqrt{\text{e}}}\sigma$/${\pi}$ 
for a good trade-off between the height of $ \phi (k) $ and the sidelobes of the photon source in Fig.2(e).
In Fig.2 (c), $A(z)$, the amplitude function  of the light field in the  crystal,  can be written as:
%%--------------------------------Eq.10-----------------------------------------
\begin{eqnarray}
A(z)=\sum\limits_{j=1}^{z/{{L}_{c}}}{g[j]\times \frac{\exp [ij{{k}_{0}}{{L}_{c}}](\exp [-i{{k}_{0}}{{L}_{c}}]-1)}{{{k}_{0}}}}.\notag\\\label{eq10}
\end{eqnarray}
In this paper, we choose an LN crystal with a length of 30 mm and the selected domain width $L_c = \Lambda/2 $. The poling distribution $g(z)$ of the crystal can be obtained by comparing $A_{\text{target}}(z) $ and $A(z)$ and fitting with the domain sequence arrangement algorithm \cite{Graffitti2017QST}, as shown in Fig.2 (d).

Figure 2 (f) is a more detailed poling distribution graph inside the 30 mm crystal.
In Fig.2 (e), according to $g(z)$, one can calculate the PMF of the PPLN crystal:
%
%%--------------------------------Eq.11-----------------------------------------
\begin{eqnarray} \label{eq11}
\phi (k)&=&\text{Abs}(\sum\limits_{j=1}^{L/{{L}_{_{c}}}}{g[j]\times \frac{i\exp [ijk{{L}_{c}}](\exp [-ik{{L}_{c}}]-1)}{k}})\notag
\\&& \times (1-2(\text{UnitStep}[k+\Omega -{{k}_{0}}]\notag
\\&& -\text{UnitStep}[k-\Omega -{{k}_{0}}])),
\end{eqnarray}
where $\Omega$ = $1.457\times {{10}^{-7}}$ rad/nm.
Finally, as shown in Fig.2 (g-i), the JSA of the designed H-G three-mode entangled state can be obtained by multiplying the PMF and the PEF.

\subsection{Preparation of comb-like entangled state for MIR wavelengths}

In this section, we design the  ten-mode comb-like entangled state in Fig.1(c1).
%---------------------------------Fig3------------------------------------------
\begin{figure*}[!tbp]
\centerline{\includegraphics[width=14cm]{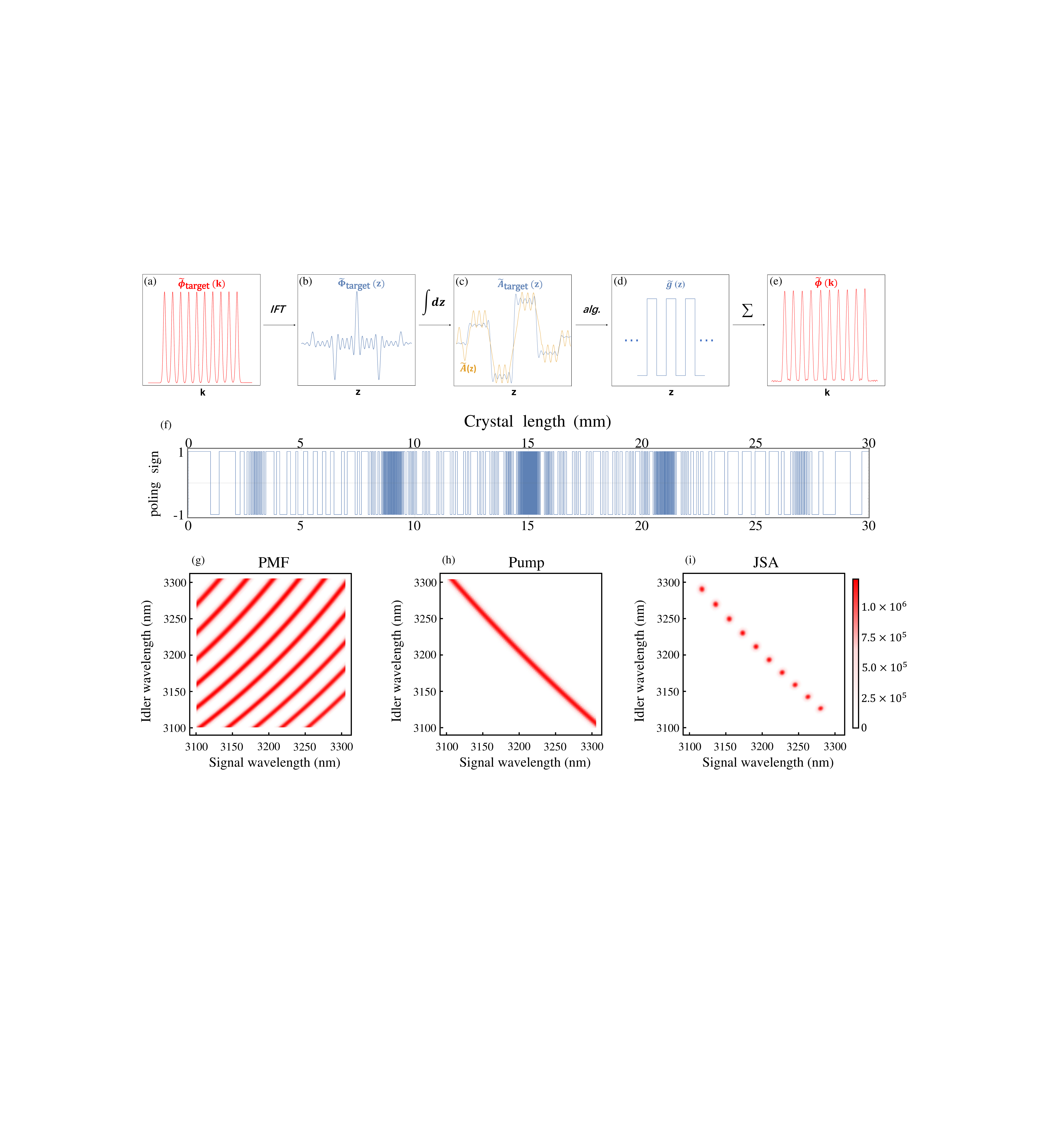}}
\centerline{\parbox[c]{13cm}{\footnotesize Fig.3. \ \ (a-e) is the design principle for a comb-like  ten-mode PMF. (f) is the designed poling distribution of the CPLN. (g-i) is the PMF, PEF, and JSA.}} \label{fig3}
\end{figure*}
As shown in Fig.3 (a), for the comb-like ten-mode entangled state, the designed PMF is:

%-------------------------------Eq.12---------------------------
\begin{eqnarray}
\widetilde{\phi}_{\text{target}}(k) &=&  \sum\limits_{n=0}^{4}(\exp[-\frac{{{\xi}^{2}}{{(k-{{k}_{0}}-(\frac{1}{2}+n)\widetilde \sigma)}^{2}}}{2}] \notag 
\\ && +\exp[-\frac{{{\xi}^{2}}{{(k-{{k}_{0}}+(\frac{1}{2}+n)\widetilde \sigma)}^{2}}}{2}]).
\end{eqnarray}
In Eq.(12),$\widetilde \sigma$ determines the width between the peaks, $\xi$ determines the width of each peak, here we take $\widetilde\sigma$ = ${k}_{0}/400 $, $\xi$ = $L/4.5$. The PMF $\widetilde{\Phi}_{\text{target}}(z)$ in the spatial domain can be obtained by IFT of Eq.(12):
%
%------------------------------Eq.13--——-----------------------------------------
\begin{equation} \label{eq13}
\widetilde{\Phi}_{\text{target}}(z)=\frac{2}{\xi }\exp [i{{k}_{0}}z-\frac{{{z}^{2}}}{2{{\xi }^{2}}}]\sum\limits_{n=0}^{4}{(\cos (2n+1)\frac{\widetilde\sigma}{2}z)}.
\end{equation}
The term $\frac{2}{\xi }\exp[-\frac{{{z}^{2}}}{2{{\xi }^{2}}}]\sum\limits_{n=0}^{4}{(\cos (2n+1)\frac{\widetilde\sigma}{2}z)}$  in Eq.(13) is first shifted by $L/2$ and then integrated to obtain the amplitude function of light field $\widetilde{A}_{\text{target}}(z)$:
%
%------------------------------Eq.14----------------------------------------------
\begin{widetext}
%\begin{equation}
\begin{eqnarray}
\widetilde{A}_{\text{target}}(z)&=& \widetilde C  \int\limits_{0}^{\text{z}}{\frac{2}{\xi }}\exp[-\frac{{{(z^{'}-\frac{L}{2})}^{2}}}{2{{\xi }^{2}}}]\sum\limits_{n=0}^{4}{(\cos (2n+1)\frac{\widetilde\sigma }{2}(z^{'}-\frac{L}{2}))}dz^{'}\notag
\\&=& \widetilde C \sum\limits_{n=0}^{4}({\sqrt{\frac{\pi }{2}}\exp [-\frac{{{(2n+1)}^{2}}}{8}{{\xi }^{2}}{{
\widetilde\sigma}^{2}}]}(-2+\text{erf}[\frac{L-(2n+1)i{{\xi }^{2}}\widetilde\sigma}{2\sqrt{2}\xi }]
+\text{erf}[\frac{L+(2n+1)i{{\xi }^{2}}
\widetilde\sigma}{2\sqrt{2}\xi }] \notag
\\&& +\text{erfc}[\frac{L-2z-(2n+1)i{{\xi }^{2}}\widetilde\sigma}{2\sqrt{2}\xi }]+\text{erfc}[\frac{L-2z+(2n+1)i{{\xi }^{2}}\widetilde\sigma}{2\sqrt{2}\xi }])).
\end{eqnarray}
%\end{equation}
\end{widetext}
%---------------------------------------------------------------------------------
In Eq.(15), we choose $ \widetilde C$ =  $\frac{5\times {{10}^{4}}L
\widetilde\sigma} {\pi}$.
As shown in Fig.3 (c), the amplitude of  the crystal light field $\widetilde{A}(z)$ can be written as:
\begin{eqnarray}
\widetilde{A}(z)=\sum\limits_{j=1}^{z/{{L}_{c}}}{\widetilde{g}[j]\times \frac{\exp [ij{{k}_{0}}{{L}_{c}}](\exp [-i{{k}_{0}}{{L}_{c}}]-1)}{{{k}_{0}}}}.\notag\\\label{eq15}
\end{eqnarray}

By comparing $ \widetilde{A}_{\text{target}}(z) $ and $\widetilde{A}(z)$ using the domain sequence arrangement algorithm  \cite{Graffitti2017QST},   we can obtain the poling distribution $\widetilde{g}(z)$ of the crystal, as shown in Fig.3 (d) and Fig.3 (f).

\begin{eqnarray} \label{eq16}
\widetilde \phi (k)&=&\text{Abs}(\sum\limits_{j=1}^{L/{{L}_{_{c}}}}{\widetilde g[j]\times \frac{i\exp [ijk{{L}_{c}}](\exp [-ik{{L}_{c}}]-1)}{k}}).\notag\\
\end{eqnarray}

The PMF of an LN crystal with a custom poling  shown in Fig.3 (e) can be calculated using Eq.(16).
Finally, as shown in Fig.3 (g-i), we multiply PMF and PEF and obtain the JSA of the designed ten-mode entangled state. 
In Fig.2(h) and Fig.3(h), the pump has a Gaussian shape with a center wavelength of 1603.8 nm and a full width at half maxima (FWHM) of 2.50 nm (corresponding to 1.51 ps in time domain).

\section{Discussion}

%\begin{table}[b]%The best place to locate the table environment is directly after its first reference in text
%\caption{\label{tab:table1}%
%Estimated pair generation rate for LN, PPLN,  CPLN \#1 (Three-mode %H-G entangled state) and CPLN \#2 (Ten-mode comb-like
%entangled state) in a type-II degenerate SPDC\footnote{In this %calculation, we use the following parameters: ${\lambda_p}=1603.8$ %$nm$ , ${n_p}=2.2026$, 
%${n_1}=2.1302$, ${n_2}=2.0612$, ${n_{g1}}=2.3276$, ${n_{g2}}=2.2335$, %${\sigma_{p}}=50$ $um$, ${\sigma_{1}}=50$ $um$, ${d_{eff}}=-3.26$ %$pm/V$, $L=30$ $mm$, $P=1$ $mW$ and ${E_p^0}=4666.6$ $V/m$.} 
%}
%\begin{ruledtabular}
%\begin{tabular}{lcdr}
%Crystal & LN &
%\multicolumn{1}{c}{\textrm{PPLN}} &
%\multicolumn{1}{c}{\textrm{CPLN$#1$\footnote{b}}} &
%\multicolumn{1}{c}{\textrm{CPLN$#2$\footnote{c}}}\\
%\hline
%${\overline \chi}(z)$ & ${1, 1, ...}$ & \mbox{${1, -1, ...}$} & %\mbox{$g(z)$} & \mbox{$\widetilde g(z)$}\\
%$\text{Rate} [s^{-1} mW^{-1}]$ & 0 & \mbox{6448} & \mbox{875} & %\mbox{371} \\
%\end{tabular}
%\end{ruledtabular}
%\end{table}

\subsection{Estimation of the photon pair rate}
It is important  to predict the pair generation rate of the designed crystals \cite{Schneeloch2019,Gong2011}. According to references \cite{Schneeloch2019}, the single-mode (spatial Gaussian mode) pair generation rate is given by:
%
%------------------------------eq.17----------------------------------------------
\begin{widetext}
\begin{eqnarray}
 {{R}_{SM}}=\frac{P}{8{{\varepsilon }_{0}}{{\pi }^{2}}{{c}^{3}}}\frac{{{n}_{g1}}{{n}_{g2}}}{n_{1}^{2}n_{2}^{2}{{n}_{p}}}{{\left|\frac{{{\sigma}_{p}}}{\sigma_{1}^{2}+2\sigma_{p}^{2}} \right|}^{2}}{{(4{{d}_{eff}})}^{2}}\int_{0}^{{{\omega }_{p}}}{{{\omega }_{s}}({{\omega }_{p}}-{{\omega }_{s}})}{{\left| \int_{0}^{L}{\overline{\chi }(z){{e}^{-ikz}}dz} \right|}^{2}}d{{\omega }_{s}}, \label{eq17}
\end{eqnarray}
\end{widetext}

where $\varepsilon_{0}$ is the vacuum permittivity, $P$ is pump power, and $P = c{\varepsilon _0}{n_p}\pi \sigma _p^2{\left| {E_p^0} \right|^2}$. $E_p^0$ is the electrical field amplitude of the pump beam, $c$ is the speed of light in vacuum, $n_{g1}(n_{g2})$ is the group index of the signal (idler) photon.  $n_{1(2,p)}$ is the the refractive index  for the signal (idler, pump) photon, ${\sigma _p}$ and ${ \sigma _1}$ refer to the beam width of the pump and the biphoton. The effective nonlinearity  ${{d}_{eff}}\equiv \frac{\chi _{eff}^{(2)}}{2}$, and $\chi_{eff}^{(2)}$ is the second order effective nonlinear susceptibility. ${\omega _s}({\omega _p})$ is the frequency of the signal (pump) beam, and ${\overline \chi}(z)$ is the poling profile in the crystal.

\begin{table*}[b]
\caption{\label{tab:table4}%
Estimated pair generation rate for LN, PPLN,  CPLN\#1 (Three-mode H-G entangled state) and CPLN\#2 (Ten-mode comb-like
entangled state) in a type-II degenerate SPDC\footnote{In this calculation, we use the following parameters: ${\lambda_p}=1603.8$ $nm$ , ${n_p}=2.2026$, 
${n_1}=2.1302$, ${n_2}=2.0612$, ${n_{g1}}=2.3276$, ${n_{g2}}=2.2335$, ${\sigma_{p}}=50$ $um$, ${\sigma_{1}}=50$ $um$, ${d_{eff}}=-3.26$ $pm/V$, $L=30$ $mm$, $P=1$ $mW$ and ${E_p^0}=4666.6$ $V/m$.}}
\begin{ruledtabular}
\begin{tabular}{ccddd}
Crystal&LN&
\multicolumn{1}{c}{\textrm{PPLN}}&
\multicolumn{1}{c}{\textrm{CPLN\#1}}&
\multicolumn{1}{c}{\textrm{CPLN\#2}} \\
\hline
${\overline \chi}(z)$&\{1, 1, ...\}&\{1, -1, ...\}
&\mbox{$g(z)$}&\mbox{$\widetilde g(z)$}\\
$\text{Rate} [s^{-1} mW^{-1}]$ &0& 6448 & 875& 371 \\
\end{tabular}
\end{ruledtabular}
\end{table*}

$k$ can be expressed using angular frequency as the variables:
%---------------------Eq.(18)--------------------------------------
\begin{eqnarray}
 k=\frac{{{\omega}_{p}}{{n}_{p}}({{\omega}_{p}})}{c}-\frac{{{\omega}_{s}}{{n}_{s}}({{\omega }_{s}})}{c}-\frac{({{\omega }_{p}}-{{\omega }_{s}}){{n}_{i}}({{\omega }_{p}}-{{\omega }_{s}})}{c}.\notag\\ \label{eq18}
\end{eqnarray}
$d_{eff}$ can be calculated using the Miller's rule:  the second order susceptibility  $\chi _{eff}^{(2)}({{\omega }_{p}},{{\omega }_{s}},{{\omega }_{i}})$ is approximately proportional to the product of the first-order susceptibilities  ${{\chi }^{(1)}}({{\omega }_{p}})$ ${{\chi }^{(1)}}({{\omega }_{s}})$ ${{\chi }^{(1)}}({{\omega }_{i}})$, and for transparent media with negligible absorption, ${{\chi }^{(1)}}(\omega )\approx n{{(\omega )}^{2}}-1$
\cite{Schneeloch2019}.
For type II SPDC  ($o\to o+e$, and 532 nm $\to$ 1064 nm + 1064 nm ) in LN crystal, $d_{eff}=d_{15}=-4.6 pm/V$ \cite{Wei2021}, so we can calculate the $d_{eff}=-3.26 pm/V$  for  SPDC of 1603.8 nm $\to$ 3207.6 nm + 3207.6 nm. 

For LN, ${\overline \chi}(z)$=\{1, 1, ...\}, for PPLN, ${\overline \chi}(z)$=\{1, -1, ...\}, for CPLN \#1 (Three-mode H-G entangled state), ${\overline \chi}(z)$=$g(z)$, and for CPLN \#2 (Ten-mode comb-like entangled state), ${\overline \chi}(z)$=$\widetilde g(z)$.
Using Eq. (17), we can estimate the pair generation rate for LN, PPLN, CPLN \#1 and CPLN \#2, as shown in Table  1.
The pair rate  are calculated to be 0 ${s^{-1}mW^{-1}}$, 6448 ${s^{-1}mW^{-1}}$, 875 ${s^{-1}mW^{-1}}$, and 371 ${s^{-1}mW^{-1}}$, respectively.
It can be concluded that CPLN \#1 and CPLN \#2 have lower rates than PPLN. There is a  trade-off between the biphoton spectral distribution and pair generation rate.

\subsection{Fabrication resolution}
It is also necessary to discuss the fabrication resolution in the experimental implementation of our designed crystal.  In Fig.4(a, b, c), we simulate the JSA using a domain width of $L_c$+100 nm, $L_c$, and $L_c$-100 nm.
Note Fig.4(b) is the same as Fig.3(i) with $L_c=7499.5$ nm.
It can be noticed that the frequency modes shift along the anti-diagonal direction by using different domain widths. Specifically, the JSA moves to the up-left (bottom-right) conner with a longer (shorter) width. 
The Schmidt number $K$, a parameter characterising  spectral  distribution, is also listed in each figure.
The $K$ is 10.0389, 10.1502, and 10.2998 for Fig.4(a, b, c).
This means a shift 100 nm in poling does not have a strong effect on the spectral correlation.

In the practical condition, the fabrication error is random.  
Based on this consideration, Fig.4(d) simulates the Schmidt number $K$ of the JSA using different fabrication resolution $R$ from  50 nm, 100 nm, 200 nm, to 400 nm.
For example, in the case of $R=100$ nm, the domain width for each domain is 
$L_c+\gamma R$, where $\gamma$ is a random number between -0.5 and 0.5.
We repeat the calculation for each $R$ for 100 times, and obtain an average Schmidt number $ \bar K$ and standard deviation $SD$.
We consider five resolutions in total: 0 nm (the ideal case), 50 nm, 100 nm, 200 nm, and 400 nm. The corresponding of $ \bar K$ and $SD$ are listed in  Fig. 4(d).
With the increase of $R$, the $ \bar K$ is almost stable, however, the $SD$ increases rapidly.  
This suggests that one need to consider the trade-off between $R$ and the $SD$  of $K$ in the fabrication process.

\begin{figure*}[!tbp]
\centerline{\includegraphics[width=13cm]{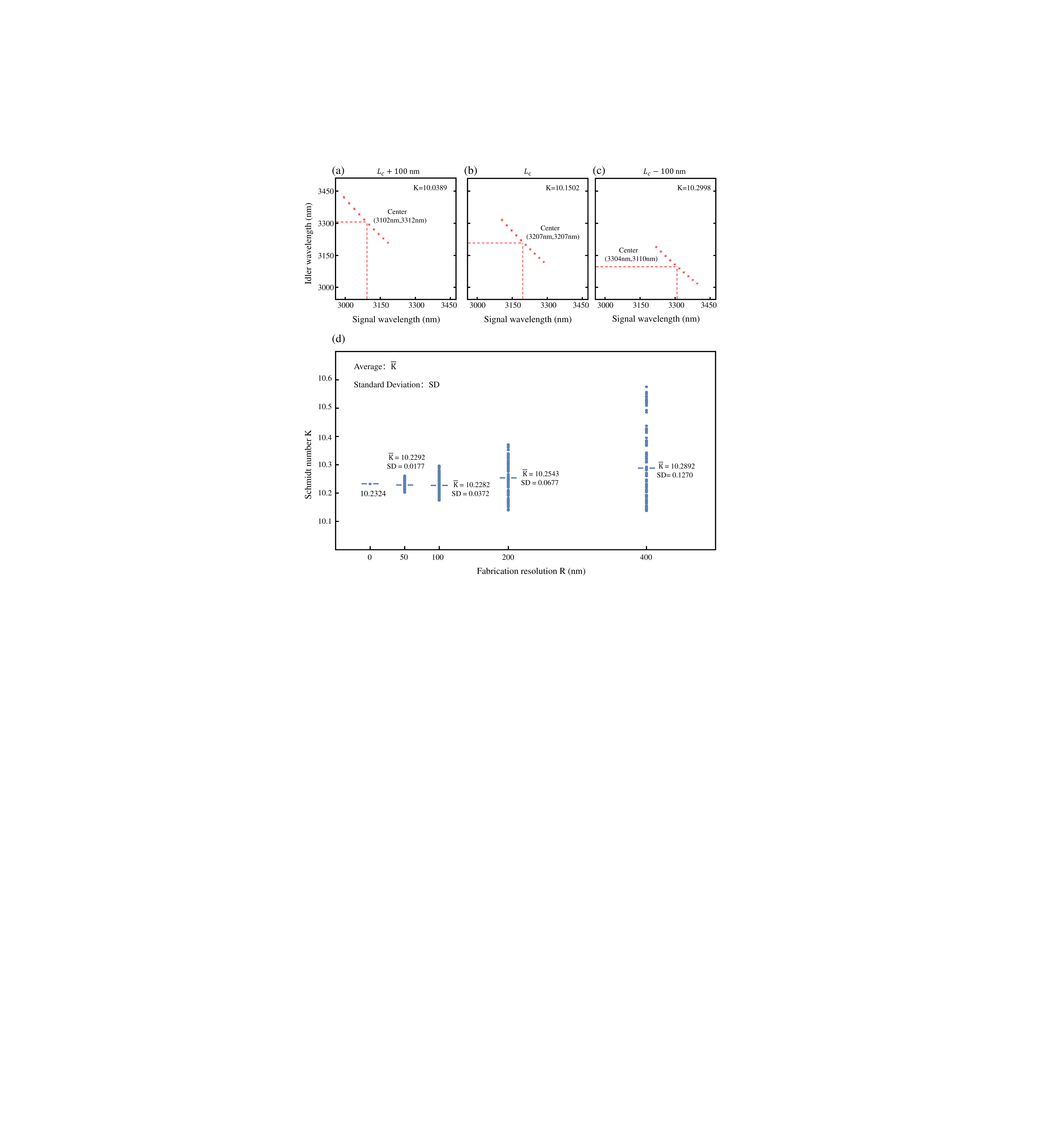}}
\centerline{\parbox[c]{13cm}{\footnotesize  Fig.4. (a-c) is the simulated JSA of three fixed domain widths: $L_c$+100 nm, $L_c$, $L_c$-100 nm. The corresponding Schmidt number $K$ is listed in each figure. (d) The distribution of Schmidt numbers $K$  calculated for 100 times using random domain width for each fabrication resolution ranges from 0 nm, 50 nm, 100 nm, 200 nm, to 400 nm.}}\label{fig4}
\end{figure*}
\subsection{Future expansion}
In Section 4, we only showed how to design the entangled state in Fig.1(a) and (c). Following a similar procedure, other JSAs in Fig.1 can also be designed by changing the PMF and PEF. To design more types of JSA, the PMF and PEF can be chosen to be arbitrary functions, e.g., the triangle function or rectangular function.

The entangled states designed in this work are time-frequency entangled states, which can be further updated to polarization-entangled state or hyper-entangled state by setting the crystal in a Sagnac-loop. In this work we only discussed the optimization of the LN crystal, other quasi-phase matched crystals are also worth investigating, e.g., the PMN-0.38PT crystal.

In this work we restrict the study in a very specific wavelength in a group-velocity matched regime. However, such frequency-shaped photons can also be produced in the non-group-velocity-matched regime.
In fact, the group-velocity matching condition only determines the ridge direction of the phase matching condition, while the custom poling on the crystal determines the comb-like structure in PMF.

\section{Conclusion}
In conclusion, we have proposed two typical time-frequency entanglement states, the H-G mode entangled state and the comb-like entangled state for MIR wavelength applications. We have also demonstrated how to design the PMF using the domain sequence arrangement algorithm.
In addition, the photon pair rate and fabrication
resolution of the designed schemes are discussed.
We hope this work can promote the study of the entangled photon source at MIR wavelengths.

\section*{Acknowledgments}
 This work was supported by the National Natural Science Foundations of China (Grant Numbers 91836102, 12074299, and 11704290).

\end{document}